\PassOptionsToPackage{table}{xcolor}
\documentclass[final,3p,times]{elsarticle}

\usepackage{algorithm}
\usepackage{algorithmic}
\usepackage{graphicx}
\usepackage{braket}
\usepackage{amsmath}
\usepackage{quantikz}
\usepackage{booktabs}
\usepackage{xcolor}
\usepackage{adjustbox}
\usepackage{listings}
\usepackage{hyperref}
\usepackage{caption}
\usepackage{multirow}

\newcommand{\QPU}{\textcolor{blue}{[QPU]}}
\newcommand{\CPU}{\textcolor{teal}{[CPU]}}

\biboptions{numbers,sort&compress}

\definecolor{neg}{RGB}{179,88,6}
\definecolor{neu}{RGB}{245,245,245}
\definecolor{pos}{RGB}{1,133,113}
\definecolor{codegray}{gray}{0.95}

\journal{Information Sciences}

\begin{document}

\begin{frontmatter}

\title{A Comparative Study of Hybrid Quantum and Classical Genetic Algorithms in Portfolio Optimization}\author[inst1]{Romeu Rossi Junior}

\ead{romeu.rossi@ufv.br}

\author[inst1]{José Augusto Miranda Nacif}

\author[inst1]{Leonardo Antônio Mendes Souza}

\author[inst1]{Marcus Henrique Soares Mendes}

\address[inst1]{Universidade Federal de Viçosa, Instituto de Ciências Exatas e Tecnológicas, Campus Florestal, 
LMG 818 Km 6, Florestal, MG 35690-000, Brazil}

\begin{abstract}
This work investigates the performance of a Hybrid Quantum Genetic Algorithm (HQGA) compared to a classical Genetic Algorithm (GA) for solving the portfolio optimization problem. Our results indicate that the HQGA converges faster to the optimal solution than its classical counterpart, while also maintaining a higher level of population diversity throughout the optimization process. In addition, the HQGA requires significantly fewer evaluations-to-solution than a brute-force approach to reach the global optimum.
\end{abstract}


\end{frontmatter}

\section{Introduction}

Quantum computing promises to revolutionize a wide range of disciplines in the coming years, and one of the most promising areas is optimization. The recent advancements in quantum hardware have brought us into what is known as the \textit{Noisy Intermediate-Scale Quantum} (NISQ) era, a term coined by John Preskill \cite{Preskill2018quantumcomputingin}. In this era, quantum devices with tens to a few hundred qubits are available, though they remain noisy, error-prone and not yet fault-tolerant. Hybrid quantum‐classical algorithms have thus emerged as one of the most practical paths toward realizing quantum advantage on NISQ hardware \cite{Callison2022hybrid}.

Among the hybrid quantum-classical approaches proposed for the NISQ era, two of the most widely studied algorithms are the Quantum Approximate Optimization Algorithm (QAOA)~\cite{Farhi2014qaoa,Hadfield2019qaoa,Egger2021warmstart,Eidenbenz2020qaoareview} and the Variational Quantum Eigensolver (VQE)~\cite{Peruzzo2014vqe,Kandala2017vqe,Tilly2022vqe,McClean2016vqa}. Both methods employ a parameterized quantum circuit whose parameters are iteratively updated by a classical optimizer: while QAOA alternates between problem-dependent and mixer Hamiltonians to approximate low-energy states of combinatorial optimization problems, VQE aims to minimize the expectation value of a Hamiltonian using a variational ansatz tailored for quantum chemistry and physics applications. In both cases, the algorithm begins from a single initial quantum state and progressively refines it by adjusting circuit parameters, steering the system toward improved solutions. This mechanism contrasts with evolutionary algorithms, which operate on a population of solutions and evolve them through selection, recombination, and mutation, progressively selecting the fittest individuals and propagating the most promising states across generations \cite{GA_class}.

Motivated by the intrinsically parallel and distributed nature of evolutionary search, several works have explored the integration of quantum-mechanical concepts into population-based optimization, leading to the development of quantum-inspired evolutionary algorithms. Quantum-inspired evolutionary algorithms emulate quantum-mechanical principles while still operating entirely on classical hardware, using abstractions such as qubits, superposition, and rotation-based update rules to guide the search \cite{Narayanan1996,Ross2019,Han2000,Han2002,Han2004}. This line of research rapidly diversified into real-coded and swarm-based variants, as well as hybrid schemes combining quantum-inspired ideas with gravitational search, differential evolution, ant colonies, immune algorithms, cuckoo search, bacterial foraging, and quantum harmonic oscillator models~\cite{Platel2008,Cruz2006,Wang2007QSwE,Soleimanpour2012,Hota2010,Honggang2009,Zhang2008,Laha2015,Gao2015,Wang2018,Wang2016,Mu2020}. Quantum-inspired mechanisms have also been incorporated into multi-objective evolutionary algorithms and successfully applied to numerous combinatorial and continuous problems, including maximum clique, 3-SAT, scheduling, and power system optimization~\cite{Kim2006,Wang2016MOEA,Dey2018,Konar2018,Das2015,Feng2008,Wu2011,Yao2012,Li2007}. Nevertheless, these approaches rely exclusively on classical hardware and Boolean logic, and thus cannot be executed directly on quantum processors~\cite{Ross2019}.

In contrast, only a few proposals aim at true quantum evolutionary algorithms, where at least part of the evolutionary cycle is implemented at the level of quantum states and gates. Some of these methods build on Grover’s quantum search algorithm to accelerate fitness evaluation and selection \cite{Grover1996,Malossini2008,Udrescu2006}, but typically delegate crossover and mutation to classical procedures and require problem-specific quantum oracles, whose compact realization may be difficult or even impractical for real-world applications.

Against this background, the Hybrid Quantum Genetic Algorithm (HQGA) proposed in \cite{Acampora2021implementing} represents a distinct approach, executing the evolutionary process directly on IBM quantum devices and introducing quantum operators that encode key mechanisms such as crossover, mutation, and elitism at the circuit level. The HQGA follows a genuinely hybrid quantum-classical workflow: the quantum processor prepares and transforms populations of chromosomes through rotation gates and entanglement, while the classical processor evaluates measured solutions and determines the next iteration. This architecture enables a complete evolutionary loop to be realized on noisy intermediate-scale quantum (NISQ) hardware, while retaining the flexibility and adaptive control typical of classical evolutionary computation.

One of the persistent challenges in genetic algorithms (GAs) is premature convergence, a phenomenon in which the population loses diversity too early and becomes trapped in a sub optimal region of the search space \cite{srinivas1994adaptive,goldberg1991comparative}. When selection pressure is high or variation operators are insufficiently disruptive, the algorithm may quickly converge to a dominant schema or allele, thereby preventing the generation of superior offspring and significantly reducing the exploratory capability of the GA \cite{srinivas1994adaptive}. In contrast, the intrinsically probabilistic nature of quantum mechanics-particularly the use of superposition and entanglement in hybrid quantum-classical schemes-offers a potential mechanism for maintaining population diversity and mitigating premature convergence in quantum-enabled evolutionary algorithms.

In recent years, portfolio optimization problems have attracted increasing attention in the field of quantum optimization. For example, in \cite{Buonaiuto2023bestpractices} the authors investigate portfolio optimization using the VQE algorithm on real quantum hardware. Shunza et al. \cite{Shunza2023discretepo} propose a discrete-asset portfolio optimization method based on quantum combinatorial optimization techniques. In \cite{Zaman2024poqa}, the authors introduce a scalable framework (PO-QA) for tuning quantum circuit parameters in portfolio optimization using QAOA and VQE. Naik et al.~\cite{Naik2025financequantum} provide an extensive survey of quantum computing applications in finance, including asset allocation. More recently, Gunjan et al. \cite{Gunjan2023quantuminspired} develop a quantum-inspired modified genetic algorithm for portfolio optimization, while Haghighi \emph{et al.} \cite{Haghighi2025eaqqa} propose the EAQGA (Quantum-Enhanced Genetic Algorithm), which demonstrates superior performance compared with a classical GA and another quantum genetic approach on a portfolio optimization dataset. Collectively, these studies demonstrate that quantum and quantum-inspired approaches are becoming viable tools for financial optimization.

In this work, we analyze the performance of the HQGA \cite{Acampora2021implementing} in the context of portfolio optimization and compare it with a classical Genetic Algorithm (GA). While HQGA has previously been shown to achieve competitive solution quality, its robustness against premature convergence, a well-known limitation of classical GAs, has received little attention. Here, we focus on this aspect and show that the HQGA maintains population diversity more effectively and converges more consistently toward the global optimum. Across multiple asset sets, population sizes, and performance metrics, our results indicate that the HQGA not only delays diversity loss but also reaches high-quality solutions with significantly fewer fitness evaluations. Taken together, these findings suggest that HQGA provides practical advantages over classical GAs, including faster convergence, reduced computational cost, and improved reliability in avoiding sub-optimal traps.

The remainder of this article is organized as follows. Section \ref{sec:quantum_basics} introduces the basic concepts of quantum computing that are relevant for understanding the hybrid algorithm discussed in this work. Section \ref{sec:portfolio} presents the portfolio optimization problem addressed in our study. Section \ref{sec:data} describes the data preparation procedures used to construct the asset sets employed in the experiments. Section \ref{sec:hqga} details the Hybrid Quantum Genetic Algorithm, including its quantum operators and hybrid execution flow. Finally, Section \ref{sec:results} reports and discusses the experimental results, with emphasis on convergence behavior, diversity preservation, and comparative performance between the HQGA and the classical GA.

\section{Basic Concepts of Quantum Computing}
\label{sec:quantum_basics}

Quantum computing relies on principles of quantum mechanics, such as superposition, interference, and entanglement, to process information in ways that classical systems cannot replicate. This section provides a brief overview of the foundational concepts required to understand the hybrid quantum-classical algorithm studied in this work.

\subsection{Qubits and Superposition}

The fundamental unit of quantum information is the qubit. Unlike a classical bit, which can only be in one of two definite states, $0$ or $1$, a qubit can exist in a coherent superposition of both basis states simultaneously. Any pure qubit state can be written as
\begin{equation}
\ket{\psi} = \alpha\ket{0} + \beta\ket{1}, \label{psi1}
\end{equation}
where $\alpha, \beta \in \mathbb{C}$ and $|\alpha|^2 + |\beta|^2 = 1$. Also, $\ket{\psi}$ is a vector in an Hilbert space, represented in equation \eqref{psi1} in th so-called computational basis $\{ \ket{0}, \ket{1} \}$ \cite{nielsen}. Equivalently, every pure qubit admits a geometric representation on the Bloch sphere of the form
\[
\ket{\psi} = \cos\left(\frac{\theta}{2}\right)\ket{0}
+ e^{i\phi}\sin\left(\frac{\theta}{2}\right)\ket{1},
\]
with real angles $\theta$ and $\phi$ parameterizing a point on the surface of a unit sphere. This parametrization highlights that a single qubit carries a continuous family of states and provides an intuitive picture of quantum manipulation: single-qubit unitary gates act as rotations of the Bloch vector. Upon a measurement in the $\{ \ket{0}, \ket{1} \}$ basis, however, the qubit collapses probabilistically to either $\ket{0}$ or $\ket{1}$ with probabilities $|\alpha|^2$ and $|\beta|^2$, reflecting the intrinsic quantum indeterminacy of superposition states \cite{nielsen}.

An essential feature of quantum superposition is that it represents a form of intrinsic probabilistic uncertainty. The state of the qubit is not merely unknown prior to measurement; rather, it is genuinely indefinite. This differs fundamentally from a classical statistical mixture. For instance, when tossing a classical coin, the result is already determined by underlying physical variables, but remains unknown to us until observation. The associated probabilities merely quantify our lack of knowledge.

In contrast, in the state $\ket{\psi} = \alpha\ket{0} + \beta\ket{1}$ we have \emph{all} the available information about the system, yet the probability that a measurement yields $0$ or $1$ remains intrinsically probabilistic. Superposition is a mathematical representation of this situation, and it has no classical counterpart: its probabilistic nature is fundamentally different from classical subjective probability, as said before, where uncertainty reflects a lack of knowledge about the underlying state. In quantum mechanics, the state contains all accessible information, and nevertheless after the measurement the system will collapse into one of the basis states. This distinction between classical uncertainty and quantum indeterminacy is central to quantum computation, as it enables interference effects and a state-space structure that have no classical analog.

\subsection{Quantum Gates as Unitary Operations}

Quantum gates transform qubit states through unitary matrices. Unlike classical logic gates, quantum gates are reversible and preserve the norm of the quantum state. A common example is the Hadamard gate \cite{nielsen, combarro},
\begin{equation}
H = \frac{1}{\sqrt{2}}
\begin{pmatrix}
1 & 1 \\
1 & -1
\end{pmatrix},
\end{equation}
which maps computational basis states into uniform superpositions:
\[
H\ket{0} = \frac{\ket{0} + \ket{1}}{\sqrt{2}}, \qquad
H\ket{1} = \frac{\ket{0} - \ket{1}}{\sqrt{2}}.
\] Another important operation is the rotation around the $y$-axis of the Bloch sphere, defined as:
\begin{equation}
R_y(\theta) =
\begin{pmatrix}
\cos(\theta/2) & -\sin(\theta/2) \\
\sin(\theta/2) & \cos(\theta/2)
\end{pmatrix}.
\end{equation}
The gate $R_y(\theta)$ continuously adjusts the probability amplitudes of a qubit and is widely used in variational and evolutionary quantum algorithms \cite{combarro}.

\subsection{Entanglement}

A defining feature of quantum systems is entanglement, a form of non-classical correlation between qubits. A two-qubit pure state $\ket{\psi}$ is separable (or factorizable) if it can be expressed as the tensor product of two single-qubit states \cite{nielsen}:
\[
\ket{\psi} = \ket{\phi_1} \otimes \ket{\phi_2}.
\]
Otherwise, the state is \emph{entangled}. For example, the Bell state
\begin{equation}
\ket{\Phi^+} = \frac{\ket{00} + \ket{11}}{\sqrt{2}}
\end{equation}
cannot be written as a product of single-qubit states and therefore represents a maximally entangled state.

Entanglement plays a central role in quantum information processing, enabling phenomena such as non-local correlations and quantum teleportation \cite{nielsen}. In the context of quantum evolutionary algorithms such as HQGA, entanglement is used to create probabilistic correlations between chromosomes, influencing the search dynamics during optimization \cite{Acampora2021implementing}.

\section{Portfolio Optimization Problem}\label{sec:portfolio}

The portfolio optimization task considered in this work follows the classical mean-variance formulation introduced by Markowitz \cite{markowitz1952portfolio}, where the goal is to maximize the expected portfolio return penalized by its risk. In this framework, the risk is quantified by the covariances (or correlations) among asset returns, since assets that tend to move together increase the overall portfolio volatility, while assets with low or negative correlations contribute to diversification and risk reduction.

The problem is expressed as a binary optimization model, in which the decision variables indicate whether each asset is included in the portfolio or not. The mathematical formulation is given by:

\begin{equation}
\max_{\mathbf{x} \in \{0,1\}^n}
\left(
\sum_{i=1}^{n} \mu_i x_i
-
\gamma \sum_{i=1}^{n} \sum_{j=1}^{n} x_i \sigma_{ij} x_j
\right), \label{markowitz}
\end{equation} where $\mu_i$ denotes the expected return of asset $i$, $\sigma_{ij}$ is the covariance between assets $i$ and $j$, and $\gamma$ is the risk-aversion coefficient that controls the trade-off between return and risk. The binary variable $x_i \in \{0,1\}$ represents the inclusion ($x_i = 1$) or exclusion ($x_i = 0$) of each asset in the optimized portfolio.

\subsection{Data Preparation}\label{sec:data}

The financial data employed in this study were collected from the Yahoo Finance database using the \texttt{yfinance} Python interface. The list of constituent companies of the S\&P 500 index was obtained from the official Wikipedia page. For each company, historical price data were retrieved over the period from October 1, 2023 to September 30, 2024, with daily frequency. The adjusted closing prices were used to compute daily returns according to:

\[
r_t = \frac{P_t}{P_{t-1}} - 1,
\]
where $P_t$ represents the adjusted price of the asset at time $t$. From these return series, two statistical quantities were computed: (i) the mean return vector $\boldsymbol{\mu}$, which represents the average expected return of each asset, and (ii) the covariance matrix $\Sigma = [\sigma_{ij}]$, which quantifies the correlations and joint variances among all pairs of assets.

These quantities were then stored for subsequent use in the optimization experiments. The mean return vector $\boldsymbol{\mu}$ serves as the reward term in the objective function, while the covariance matrix $\Sigma$ defines the quadratic penalty associated with portfolio risk.

To evaluate the robustness and convergence properties of the algorithms, multiple independent optimization instances were generated, each corresponding to a distinct subset of assets randomly sampled from the S\&P~500. In this study, five subsets, each containing nine assets, were randomly generated. This procedure allows the algorithms to be tested under diverse asset combinations while keeping the problem dimensionality manageable. Each subset therefore defines a separate binary optimization instance, characterized by its own mean-return vector and covariance matrix. Each optimization instance was encoded as a \textit{binary problem} within the HQGA framework. The decision vector $\mathbf{x} = (x_1, x_2, \dots, x_n)$ determines which assets are included in the portfolio. The fitness function directly implements the Markowitz objective, Equation \eqref{markowitz}, combining expected return and risk under the specified risk-aversion parameter $\gamma$. 

This formulation enables both the classical Genetic Algorithm (GA) and the Hybrid Quantum Genetic Algorithm (HQGA) to operate over the same discrete search space, allowing a fair and consistent comparison between the two methods.

\section{Hybrid Quantum Genetic Algorithm}\label{sec:hqga}

The Hybrid Quantum Genetic Algorithm (HQGA) was originally proposed by Giovanni Acampora and Autilia Vitiello \cite{Acampora2021implementing}. The HQGA is designed to operate on a hybrid computational architecture in which classical and quantum processors interact to perform evolutionary optimization. In this architecture, the quantum processor is responsible for implementing the evolutionary mechanisms such as population evolution, mutation, and entanglement-based crossover, while the classical processor performs the fitness evaluations of the candidate solutions.

The structure of the HQGA resembles that of classical genetic algorithms, but incorporates four additional mechanisms rooted in quantum mechanics, that significantly increase the robustness of the method against premature convergence, as will be detailed in the following sections. The first of these mechanisms concerns the representation of each individual as a quantum state. Because a quantum state in superposition can correspond to any feasible solution of the problem, the HQGA naturally exploits the intrinsic probabilistic nature of quantum mechanics to maintain a high level of diversity in the population. This quantum representation allows the algorithm to explore a substantially larger portion of the search space than classical methods, without incurring an exponential computational cost.

The second quantum mechanism incorporated into the HQGA is the quantum elitism, responsible for ensuring that the best solution found by the algorithm in each iteration is preserved in the quantum population. Unlike classical elitism, the probabilistic nature of quantum states requires a different strategy. In the HQGA, the best individual exists simultaneously in two forms: a classical bit string obtained after measurement and a quantum state whose amplitudes encode the probability of collapsing to that configuration. Quantum elitism reconstructs or reinforces this quantum state at every iteration so that the best solution remains available as a reliable reference for the evolutionary operators. Three variants are employed: \textit{pure quantum elitism}, which faithfully reconstructs the previous best quantum state; \textit{quantum elitism with reinforcement}, which increases the probability of collapsing to the best classical solution; and \textit{deterministic quantum elitism}, which forces the quantum state to encode the best individual exactly. This mechanism guarantees stability in the evolutionary process while maintaining the quantum representation necessary for subsequent operators such as entangled crossover.

The third quantum mechanism incorporated into the HQGA is the entangled crossover, the main evolutionary operator responsible for propagating useful information throughout the population. Rather than copying genetic material as in classical crossover, the HQGA employs quantum entanglement to establish probabilistic correlations between the qubits of the best individual and those of the remaining individuals. In practice, portions of the best chromosome are entangled with corresponding regions of the other chromosomes, such that, upon measurement, these individuals have an increased likelihood of collapsing into states that share structural similarities with the current best solution. This mechanism does not enforce exact replication; instead, it induces a stochastic influence that preserves diversity while guiding the search toward promising regions of the solution space. Consequently, the entangled crossover simultaneously enhances exploitation of high-quality solutions and maintains the exploratory behavior necessary to avoid premature convergence.

The fourth quantum mechanism of the HQGA is the \textit{$R_y$ mutation}, responsible for introducing controlled stochastic perturbations in the population and enhancing the algorithm's exploratory capability. After the entangled crossover is applied, only part of each chromosome becomes correlated with the best individual; the remaining qubits, known as free qubits, retain their previous quantum states. These free qubits provide an opportunity for exploration and are subjected to a mutation step implemented through rotations around the $y$-axis of the Bloch sphere. With a given mutation probability, the $R_y$ gate adjusts the amplitudes of these qubits, altering the likelihood that they collapse to~0 or~1 in the next measurement. Unlike classical mutation, which flips bits directly, the $R_y$ mutation modifies the underlying probability distribution of each qubit, allowing the HQGA to explore new regions of the search space while still maintaining a quantum representation consistent with the evolutionary dynamics.

The HQGA operates through a hybrid quantum classical cycle that leverages the strengths of both computational paradigms to perform evolutionary optimization. The quantum subsystem is responsible for representing and evolving the population through the four quantum mechanisms previously introduced quantum representation, quantum elitism, entangled crossover, and $R_y$ mutation, while the classical subsystem evaluates the candidate solutions. Thanks to the ability of quantum registers to encode a large number of states in superposition, the HQGA can explore many potential solutions simultaneously. After each quantum evolution step, a measurement collapses the quantum chromosomes into a classical bit strings, which are then evaluated efficiently by the classical computer. This synergy allows the algorithm to navigate the search space in a manner that would be computationally prohibitive for classical genetic algorithms alone.

The hybrid cycle begins with a quantum initialization stage, in which a quantum circuit prepares the initial population of quantum chromosomes. A measurement is then performed to obtain classical candidate solutions, whose fitness values are computed by the classical processor. The best classical individual identified in this evaluation step is used to compute the parameters required for the next quantum evolution, namely: (i) the reconstruction of the quantum states of the non-best individuals, (ii) the selection of qubits that will be entangled with the best individual, and (iii) the rotation angles for the $R_y$ mutation applied to the free qubits. The configuration of the quantum elitism mechanism, however, is defined from the best quantum state preserved in the previous iteration. Once these elements are specified, the quantum evolutionary circuit is executed to generate a new quantum population. The algorithm then iterates between quantum evolution and classical evaluation until the stopping criterion is met.

The final output is the best-performing solution identified throughout the optimization process. Through this hybrid workflow, the HQGA takes advantage of the quantum computer’s ability to explore large portions of the search space in superposition, while the classical computer provides reliable fitness evaluation and convergence control. The structure of the algorithm is summarized in Algorithm~\ref{alg:HQGA}.

\begin{algorithmic}[1]
\captionsetup{font=normalsize, justification=centering}
\captionof{algorithm}{Hybrid Quantum Genetic Algorithm (HQGA). 
The steps executed on a quantum processor are indicated by \textcolor{blue}{[QPU]}, 
while the classical computations are marked with \textcolor{teal}{[CPU]}.}\label{alg:HQGA} 
{\centering
\rule{0.9\linewidth}{0.5pt}\par
}
\REQUIRE Optimization problem, population size $N$, mutation probability $p_m$, max iterations $T$
\ENSURE Best solution $x^\star$

\STATE \textbf{Initialization:}
\STATE \QPU\ Prepare initial quantum population 
       $\{ |\psi_1^{(0)}\rangle, \ldots, |\psi_N^{(0)}\rangle \}$
\STATE \QPU\ Measure each quantum chromosome to obtain classical bit strings 
       $\{ x_1^{(0)}, \ldots, x_N^{(0)} \}$
\STATE \CPU\ Evaluate fitness of all individuals and identify best solution $x^\star_{(0)}$
\STATE \QPU\ Reconstruct best quantum state $|\psi^\star_{(0)}\rangle$ according to elitism strategy

\FOR{$t = 1$ to $T$}

    \STATE \textbf{Classical Evaluation:}
    \STATE \QPU\ Measure all quantum chromosomes 
           $|\psi_i^{(t-1)}\rangle \rightarrow x_i^{(t-1)}$
    \STATE \CPU\ Evaluate fitness of each $x_i^{(t-1)}$
    \STATE \CPU\ Identify best classical individual $x^\star_{(t)}$
    \IF{$f(x^\star_{(t)}) < f(x^\star_{(t-1)})$}
        \STATE \CPU\ Update stored best solution $x^\star \leftarrow x^\star_{(t)}$
    \ENDIF

    \STATE \textbf{Quantum Elitism:}
    \STATE \QPU\ Reconstruct or reinforce the quantum state $|\psi^\star_{(t)}\rangle$ encoding $x^\star$

    \STATE \textbf{Entangled Crossover:}
    \FOR{each non-best chromosome $|\psi_i^{(t-1)}\rangle$}
        \STATE \QPU\ Select qubits to be entangled with best individual
        \STATE \QPU\ Apply entangling operator $U_{\mathrm{ent}}$ 
                     to correlate $|\psi^\star_{(t)}\rangle$ and $|\psi_i^{(t-1)}\rangle$
        \STATE \QPU\ Produce updated chromosome $\tilde{\psi}_i^{(t)}$
    \ENDFOR

    \STATE \textbf{$R_y$ Mutation:}
    \FOR{each chromosome $|\tilde{\psi}_i^{(t)}\rangle$}
        \STATE \QPU\ Identify free (non-entangled) qubits
        \FOR{each free qubit $q$}
            \STATE \QPU\ With probability $p_m$, apply rotation $R_y(\theta_q)$
        \ENDFOR
        \STATE \QPU\ Obtain mutated state $|\psi_i^{(t)}\rangle$
    \ENDFOR

\ENDFOR

\STATE \CPU\ \textbf{Return} the best solution $x^\star$ found during the process

{\centering
\rule{0.9\linewidth}{0.5pt}\par
}

\end{algorithmic}

\section{Results and Discussion}\label{sec:results}

The goal of this study is to assess whether the HQGA exhibits greater resistance to premature convergence compared to a classical Genetic Algorithm (GA). Since premature convergence is often associated with a rapid loss of population diversity, an algorithm that maintains a broader exploration capability is less likely to become trapped in local optima. The intrinsic randomness of quantum states employed in the HQGA, together with the quantum operators acting on superpositions of solutions, suggests that the algorithm may preserve diversity more effectively than its classical counterpart. To investigate this hypothesis, we analyze the diversity dynamics of both algorithms throughout the optimization process.

To quantify the degree of population diversity, we adopt the metric proposed by Srinivas and Patnaik~\cite{srinivas1994adaptive}, defined as the difference between the maximum fitness $f_{\max}$ and the average fitness $\bar{f}$ of the population. This quantity serves as an effective indicator of diversity: when the population becomes homogeneous and converges prematurely to a local optimum, the values of its individuals become similar, causing $f_{\max} - \bar{f}$ to decrease. Conversely, larger values of $f_{\max} - \bar{f}$ indicate a population that remains more dispersed in the search space. This metric has been widely used to diagnose convergence behavior in genetic algorithms, and here it allows us to compare the diversification dynamics of HQGA and a classical GA. As shown in the following results, HQGA consistently maintains higher values of $f_{\max} - \bar{f}$, demonstrating its enhanced ability to preserve population diversity and avoid premature convergence.

\begin{figure}[ht]
    \centering
    \includegraphics[width=\linewidth]{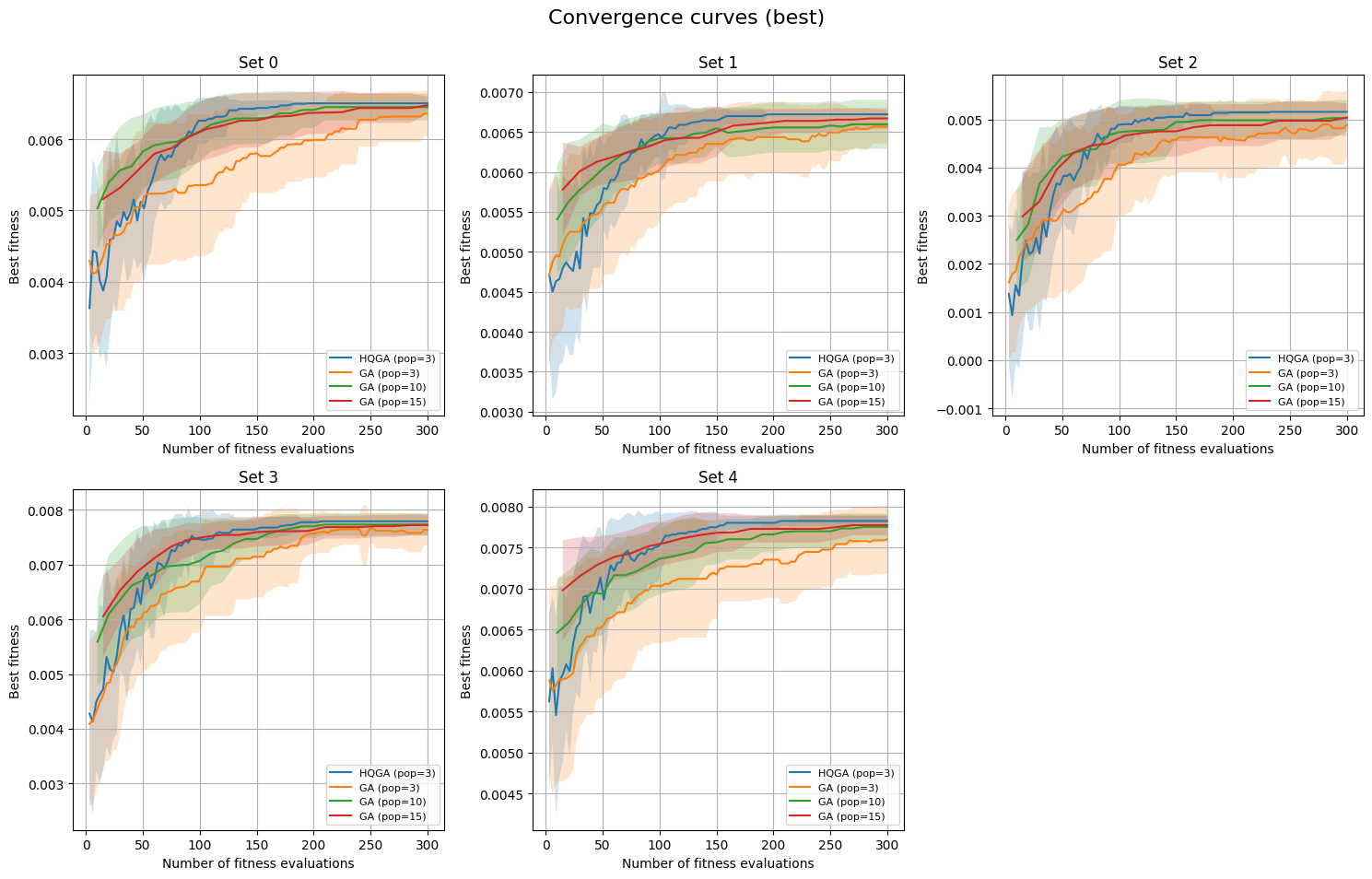}
    \caption{Convergence curves for the best fitness across all asset sets.}
    \label{fig:conv_best}
\end{figure}

\begin{figure}[ht]
    \centering
    \includegraphics[width=\linewidth]{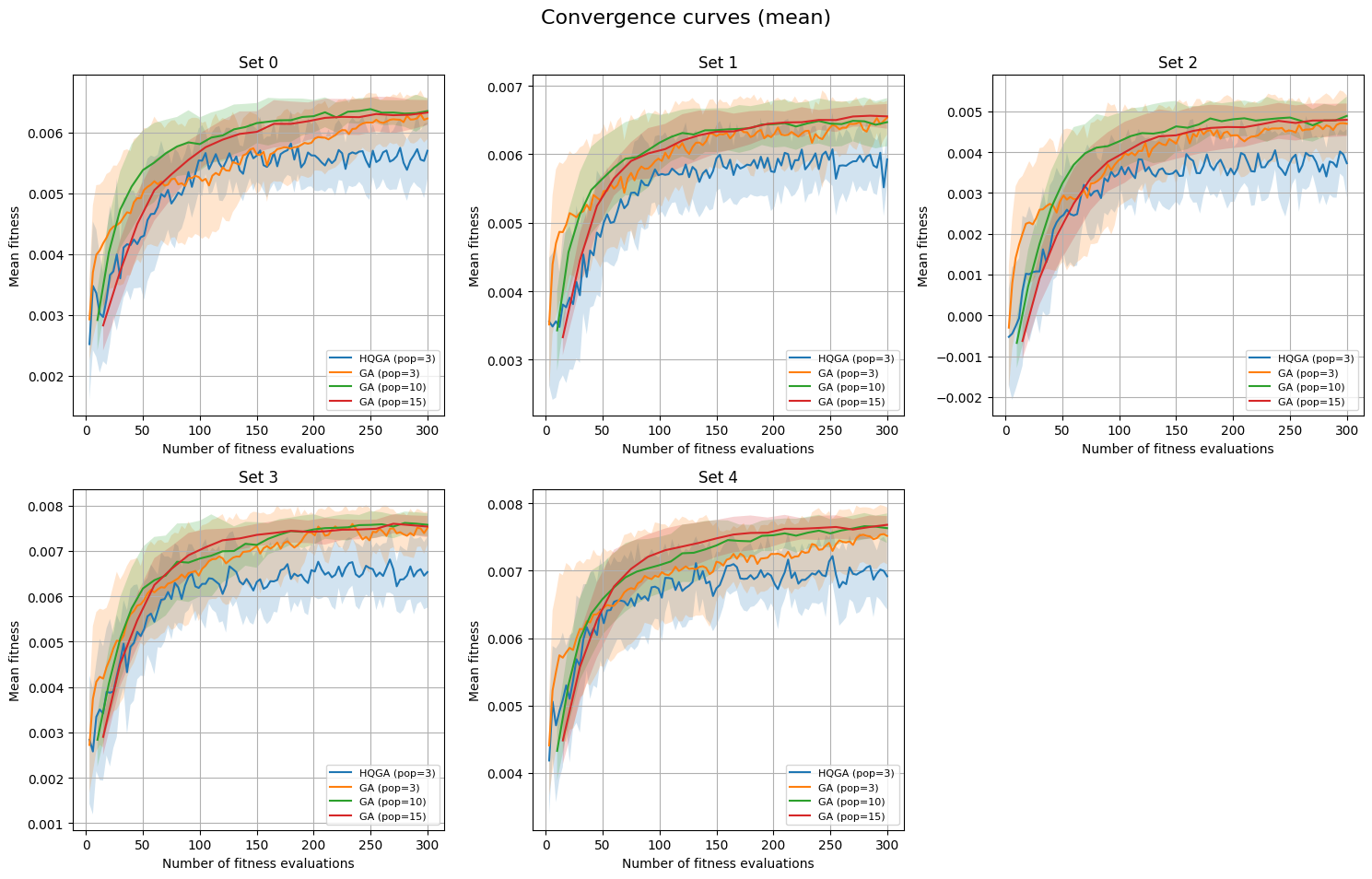}
    \caption{Convergence curves for the mean fitness across all asset sets.}
    \label{fig:conv_mean}
\end{figure}

\begin{figure}[ht]
    \centering
    \includegraphics[width=\linewidth]{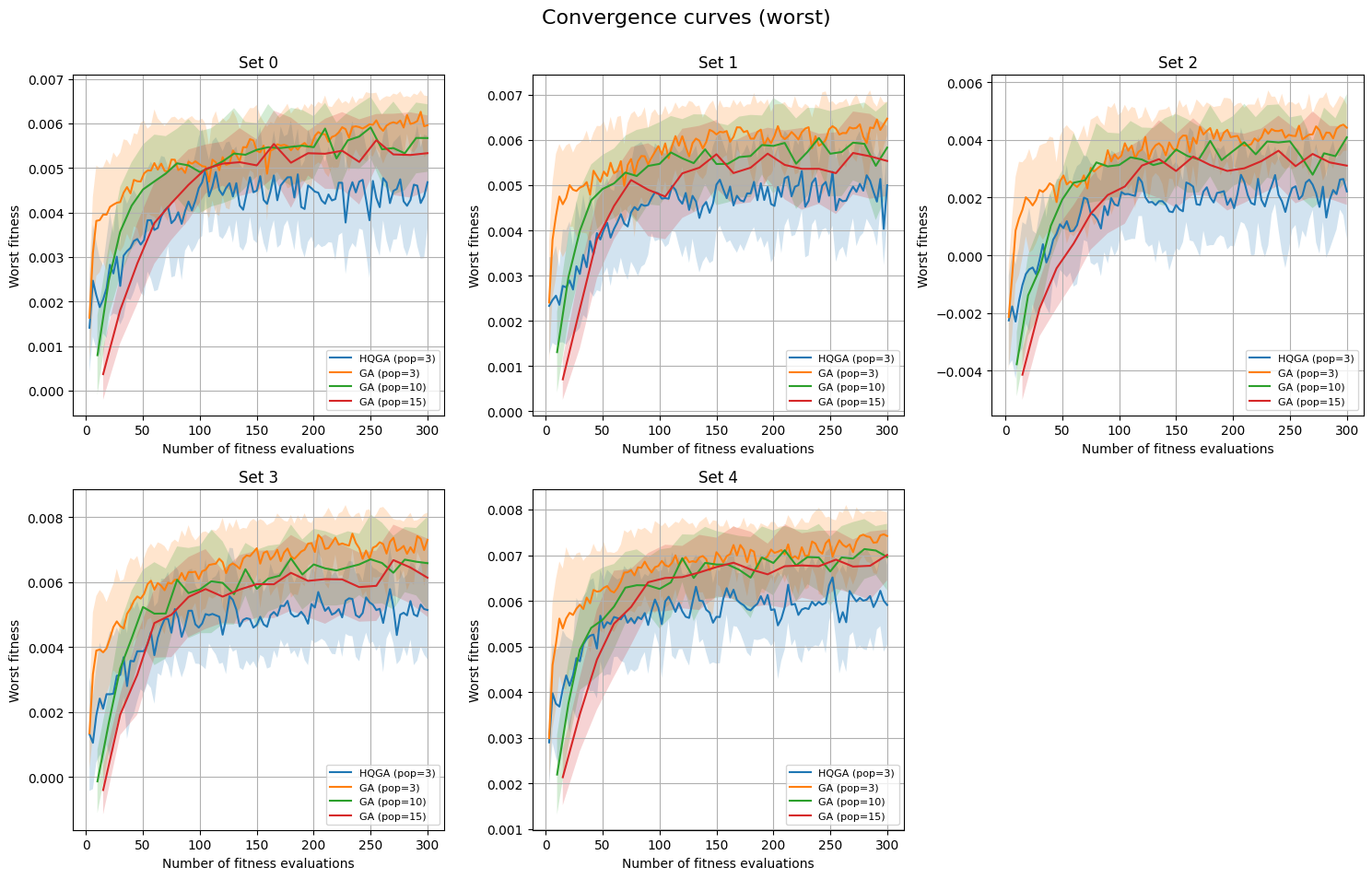}
    \caption{Convergence curves for the worst fitness across all asset sets.}
    \label{fig:conv_worst}
\end{figure}

We first present the convergence results obtained with the HQGA-based approach to our problem. We then analyze population diversity, which further supports and validates our findings. Figures~\ref{fig:conv_best}, \ref{fig:conv_mean}, and~\ref{fig:conv_worst} present the convergence curves for the HQGA and the classical GA across five independent asset sets. For each algorithm and population size, these figures show the evolution of the best (Figure~\ref{fig:conv_best}), mean (Figure~\ref{fig:conv_mean}), and worst (Figure~\ref{fig:conv_worst}) fitness values as a function of the number of fitness evaluations. This axis is particularly meaningful because the number of fitness evaluations is one of the most relevant computational-cost indicators in evolutionary algorithms, providing a hardware-independent basis for comparing convergence behaviors. The solid lines represent the average fitness behavior over all individuals sharing the same evaluation count, while the shaded regions indicate one standard deviation. Together, these plots allow us to compare the convergence speed and robustness of both algorithms across different asset sets and population sizes.

Across all asset sets, the convergence curves reveal a consistent trend in favor of the HQGA. For the best fitness values, the HQGA with a small population size (pop = 3) converges as quickly as, or even faster than, the classical GA with larger populations, highlighting the efficiency of the quantum evolutionary mechanisms. The mean-fitness curves further indicate that HQGA maintains a stable progression toward high-quality regions of the search space, while the worst-fitness plots show that the algorithm improves even its lowest-performing individuals throughout the optimization. Moreover, in all experiments the HQGA reaches the global optimum with significantly fewer fitness evaluations than the brute-force method (which in this case requires exhaustively evaluating all $2^9 = 512$ possible portfolios), demonstrating its ability to identify optimal solutions early in the search. These results show that the HQGA achieves competitive convergence performance while operating with substantially smaller populations and reduced computational cost.

While convergence curves provide insights into the optimization dynamics of each algorithm, they do not directly reveal how diversity evolves within the population over time. Since premature convergence is typically associated with a rapid loss of diversity, it is equally important to analyze how the distribution of fitness values behaves throughout the search process. To complement the convergence analysis, we therefore examine the evolution of the diversity metric $f_{\max} - \bar{f}$, previously mentioned, for both HQGA and the classical GA across all asset sets and population sizes.

\begin{figure}[ht]
    \centering
    \includegraphics[width=\linewidth]{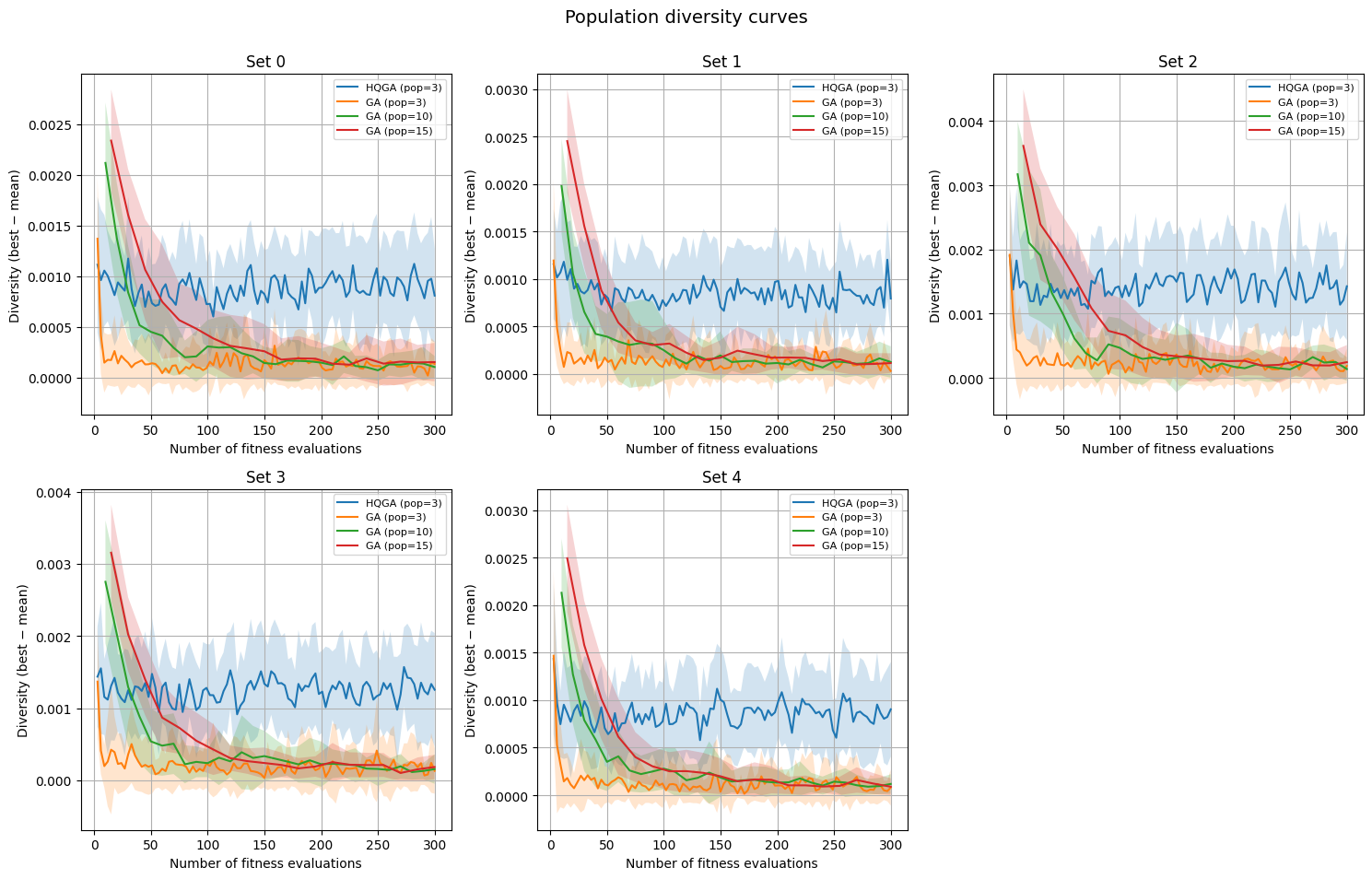}
    \caption{Population diversity curves for the HQGA and the classical GA across all asset sets. The diversity metric is defined as the difference between the best and mean fitness values ($f_{\max} - \bar{f}$), evaluated over the number of fitness evaluations.}
    \label{fig:diversity}
\end{figure}

Figure \ref{fig:diversity} displays the population diversity curves computed for all five asset sets, using the metric defined earlier. For each algorithm and population size, the diversity is evaluated at every fitness evaluation step and averaged across individuals sharing the same evaluation count. The shaded regions represent one standard deviation, providing a notion of variability in the spread of the population. One can note in Figure~\ref{fig:diversity} that, across all five asset sets, the HQGA maintains substantially higher diversity levels than the classical GA throughout the entire optimization process. For all population sizes considered, the diversity curves of the GA decrease rapidly during the first evaluations and then stabilize near zero, indicating that the classical GA quickly becomes homogeneous and loses its exploratory capability. In contrast, the HQGA consistently preserves a non-negligible diversity level, even after hundreds of fitness evaluations. This behavior is particularly evident for the smallest population size (pop = 3), where the contrast between the two algorithms is most pronounced: while the GA collapses to very low diversity almost immediately, the HQGA sustains a markedly broader distribution of fitness values over time.

These results demonstrate that the quantum mechanisms embedded in the HQGA, especially quantum superposition, entangled crossover, and $R_y$ mutation, play a crucial role in preventing the loss of population diversity. By maintaining higher values of the metric $f_{\max} - \bar{f}$, the HQGA avoids the rapid homogenization commonly observed in classical genetic algorithms and therefore reduces the likelihood of premature convergence. The sustained diversity observed across all asset sets indicates that the HQGA is able to continue exploring new regions of the search space even in later stages of the optimization, providing a more robust search dynamics and increasing the probability of escaping local optima.

\section{Conclusion}

In this work, we evaluated the performance of the Hybrid Quantum Genetic Algorithm (HQGA) in the context of portfolio optimization and compared it with a classical Genetic Algorithm (GA). Our results demonstrate that the HQGA consistently achieves competitive or superior convergence performance across multiple asset sets, even when operating with significantly smaller population sizes. The algorithm rapidly approaches high-quality solutions and, in all experiments, reaches the global optimum with substantially fewer fitness evaluations than the brute-force approach, which in this case requires exhaustively evaluating all $2^9 = 512$ possible portfolios, highlighting the efficiency of the HQGA in reducing computational cost.

A key advantage of the HQGA lies in its ability to preserve population diversity throughout the optimization process. By leveraging quantum specific mechanisms such as superposition-based representation, quantum elitism, entangled crossover, and $R_y$ mutation, the HQGA maintains a broader distribution of fitness values, avoiding premature convergence and enabling sustained exploration of the search space. In contrast, the classical GA quickly loses diversity, especially for small populations, which increases the risk of becoming trapped in local optima.

Overall, the empirical evidence indicates that hybrid quantum-classical evolutionary strategies can offer meaningful benefits over purely classical methods, particularly for combinatorial optimization problems requiring both effective exploration and rapid convergence. As quantum hardware continues to advance, future work may involve scaling the HQGA to larger portfolios, exploring deeper circuit designs for evolutionary operators, and investigating how quantum noise and decoherence influence the algorithm’s behavior in realistic quantum environments. These investigations will further clarify the potential role of hybrid quantum algorithms in practical financial optimization tasks.

\section*{Acknowledgements}

This study was partially funded by the Coordenação de Aperfeiçoamento de Pessoal de Nível Superior (CAPES), Fapemig, and CNPq. LAMS acknowledges FAPEMIG/MCI-Confap (APQ-06255-25).

\bibliographystyle{elsarticle-num}

\begin{thebibliography}{99}


\bibitem{Preskill2018quantumcomputingin}
Preskill, J. (2018).
\textit{Quantum computing in the NISQ era and beyond}.
Quantum, 2(79).


\bibitem{Callison2022hybrid}
Callison, A., \& Chancellor, N. (2022).
\textit{Hybrid quantum-classical algorithms in the noisy intermediate-scale quantum era and beyond}.
Physical Review A, 106(1), 010101.



\bibitem{Farhi2014qaoa}
Farhi, E., Goldstone, J., \& Gutmann, S. (2014).
\textit{A quantum approximate optimization algorithm}.
arXiv:1411.4028.

\bibitem{Hadfield2019qaoa}
Hadfield, S., Wang, Z., O'Gorman, B., Rieffel, E. G., Venturelli, D., \& Biswas, R. (2019).
\textit{From the quantum approximate optimization algorithm to a quantum alternating operator ansatz}.
Algorithms, 12(2), 34.

\bibitem{Egger2021warmstart}
Egger, D. J., Mareček, J., \& Woerner, S. (2021).
\textit{Warm-starting quantum optimization}.
Quantum, 5, 479.

\bibitem{Eidenbenz2020qaoareview}
Eidenbenz, S. (2020).
\textit{QAOA for Max-Cut: A review}.
arXiv:2005.01088.



\bibitem{Peruzzo2014vqe}
Peruzzo, A., McClean, J., Shadbolt, P., Yung, M.-H., Zhou, X. Q., Love, P., ... \& O'Brien, J. L. (2014).
\textit{A variational eigenvalue solver on a photonic quantum processor}.
Nature Communications, 5, 4213.

\bibitem{Kandala2017vqe}
Kandala, A., Mezzacapo, A., Temme, K., et al. (2017).
\textit{Hardware-efficient variational quantum eigensolver for small molecules and quantum magnets}.
Nature, 549(7671), 242-246.

\bibitem{Tilly2022vqe}
Tilly, J., Chen, H., Cao, S., et al. (2022).
\textit{The variational quantum eigensolver: A review of methods and applications}.
Physics Reports, 986, 1-128.

\bibitem{McClean2016vqa}
McClean, J. R., Romero, J., Babbush, R., \& Aspuru-Guzik, A. (2016).
\textit{The theory of variational hybrid quantum-classical algorithms}.
New Journal of Physics, 18(2), 023023.

\bibitem{GA_class}
Eiben, A. E., Smith, J. E. (2015). Introduction to Evolutionary Computing (2nd ed.). Springer.


\bibitem{Narayanan1996}
Narayanan, A., \& Moore, M. (1996).
\textit{Quantum-inspired genetic algorithms}.
Proceedings of IEEE International Conference on Evolutionary Computation, 61-66.

\bibitem{Ross2019}
Ross, O.H.M. (2019).
\textit{A review of quantum-inspired metaheuristics: going from classical computers to real quantum computers}.
IEEE Access, 8, 814-838.

\bibitem{Han2000}
Han, K.-H., \& Kim, J.-H. (2000).
\textit{Genetic quantum algorithm and its application to combinatorial optimization problem}.
Proceedings of the 2000 Congress on Evolutionary Computation, Vol. 2, 1354-1360.

\bibitem{Han2002}
Han, K.-H., \& Kim, J.-H. (2002).
\textit{Quantum-inspired evolutionary algorithm for a class of combinatorial optimization}.
IEEE Transactions on Evolutionary Computation, 6(6), 580-593.

\bibitem{Han2004}
Han, K.-H., \& Kim, J.-H. (2004).
\textit{Quantum-inspired evolutionary algorithms with a new termination criterion, $\epsilon$-gate, and two-phase scheme}.
IEEE Transactions on Evolutionary Computation, 8(2), 156-169.

\bibitem{Platel2008}
Platel, M.D., Schliebs, S., \& Kasabov, N. (2008).
\textit{Quantum-inspired evolutionary algorithm: a multimodel EDA}.
IEEE Transactions on Evolutionary Computation, 13(6), 1218-1232.

\bibitem{Cruz2006}
Abs da Cruz, A.V., Vellasco, M.M.B.R., \& Pacheco, M.A.C. (2006).
\textit{Quantum-inspired evolutionary algorithm for numerical optimization}.
2006 IEEE International Conference on Evolutionary Computation, 2630-2637.

\bibitem{Wang2007QSwE}
Wang, Y., Feng, X.-Y., Huang, Y.-X., Pu, D.-B., Zhou, W.-G., Liang, Y.-C., \& Zhou, C.-G. (2007).
\textit{A novel quantum swarm evolutionary algorithm and its applications}.
Neurocomputing, 70(4), 633-640.

\bibitem{Soleimanpour2012}
Soleimanpour-Moghadam, M., \& Nezamabadi-Pour, H. (2012).
\textit{An improved quantum behaved gravitational search algorithm}.
20th Iranian Conference on Electrical Engineering (ICEE2012), 711-715.

\bibitem{Hota2010}
Hota, A.R., \& Pat, A. (2010).
\textit{An adaptive quantum-inspired differential evolution algorithm for 0-1 knapsack problem}.
2010 Second World Congress on Nature and Biologically Inspired Computing, 703-708.

\bibitem{Honggang2009}
Honggang, W., Liang, M., Huizhen, Z., \& Gaoya, L. (2009).
\textit{Quantum-inspired ant algorithm for knapsack problems}.
Journal of Systems Engineering and Electronics, 20(5), 1012-1016.

\bibitem{Zhang2008}
Zhang, X. (2008).
\textit{Quantum-inspired immune evolutionary algorithm}.
International Seminar on Business and Information Management, Vol. 1, 323-325.

\bibitem{Laha2015}
Laha, S. (2015).
\textit{A quantum-inspired cuckoo search algorithm for the travelling salesman problem}.
2015 International Conference on Computing, Communication and Security, 1-6.

\bibitem{Gao2015}
Gao, H., \& Li, C. (2015).
\textit{Quantum-inspired bacterial foraging algorithm for parameter adjustment in green cognitive radio}.
Journal of Systems Engineering and Electronics, 26(5), 897-907.

\bibitem{Wang2018}
Wang, P., Ye, X., Li, B., \& Cheng, K. (2018).
\textit{Multi-scale quantum harmonic oscillator algorithm for global numerical optimization}.
Applied Soft Computing, 69, 655-670.

\bibitem{Wang2016}
Wang, P., \& Huang, Y. (2016).
\textit{MQHOA algorithm with energy level stabilizing process}.
Journal of Communications, 37(7), 79-86.

\bibitem{Mu2020}
Mu, L., Wang, P., \& Xin, G. (2020).
\textit{Quantum-inspired algorithm with fitness landscape approximation in reduced dimensional spaces for numerical function optimization}.
Information Sciences, 527, 253-278.

\bibitem{Kim2006}
Kim, Y., Kim, J.-H., \& Han, K.-H. (2006).
\textit{Quantum-inspired multiobjective evolutionary algorithm for multiobjective 0/1 knapsack problems}.
2006 IEEE International Conference on Evolutionary Computation, 2601-2606.

\bibitem{Wang2016MOEA}
Wang, Y., Li, Y., \& Jiao, L. (2016).
\textit{Quantum-inspired multi-objective evolutionary algorithm based on decomposition}.
Soft Computing, 20(8), 3257-3272.

\bibitem{Dey2018}
Dey, S., Bhattacharyya, S., \& Maulik, U. (2018).
\textit{Quantum inspired nondominated sorting based multi-objective GA for multi-level image thresholding}.
In Hybrid Metaheuristics: Research and Applications, 141-170.

\bibitem{Konar2018}
Konar, D., Sharma, K., Sarogi, V., \& Bhattacharyya, S. (2018).
\textit{A multi-objective quantum-inspired genetic algorithm (MO-QIGA) for real-time task scheduling}.
Procedia Computer Science, 131, 591-599.

\bibitem{Das2015}
Das, P.P., \& Khan, M.H. (2015).
\textit{Solving maximum clique problem using a novel quantum-inspired evolutionary algorithm}.
2015 International Conference on Electrical Engineering and Information Communication Technology, 1-6.

\bibitem{Feng2008}
Feng, X., Blanzieri, E., \& Liang, Y. (2008).
\textit{Improved quantum-inspired evolutionary algorithm and its application to 3-SAT problems}.
2008 International Conference on Computer Science and Software Engineering, Vol. 1, 333-336.

\bibitem{Wu2011}
Wu, X., \& Li, S. (2011).
\textit{A quantum inspired algorithm for the job shop scheduling problem}.
2011 IEEE 2nd International Conference on Computing, Control and Industrial Engineering, Vol. 2, 212-215.

\bibitem{Yao2012}
Yao, F., Dong, Z.Y., Meng, K., Xu, Z., Iu, H.H.-C., \& Wong, K.P. (2012).
\textit{Quantum-inspired particle swarm optimization for power system operations}.
IEEE Transactions on Industrial Informatics, 8(4), 880-888.

\bibitem{Li2007}
Li, B.-B., \& Wang, L. (2007).
\textit{A hybrid quantum-inspired genetic algorithm for multiobjective flow shop scheduling}.
IEEE Transactions on Systems, Man, and Cybernetics Part B, 37(3), 576-591.

\bibitem{Grover1996}
Grover, L.K. (1996).
\textit{A fast quantum mechanical algorithm for database search}.
Proceedings of the Twenty-Eighth Annual ACM Symposium on Theory of Computing, 212-219.

\bibitem{Malossini2008}
Malossini, A., Blanzieri, E., \& Calarco, T. (2008).
\textit{Quantum genetic optimization}.
IEEE Transactions on Evolutionary Computation, 12(2), 231-241.

\bibitem{Udrescu2006}
Udrescu, M., Prodan, L., \& Vlăduțiu, M. (2006).
\textit{Implementing quantum genetic algorithms: a solution based on Grover’s algorithm}.
Proceedings of the 3rd Conference on Computing Frontiers, 71-82.



\bibitem{Acampora2021implementing}
Acampora, G., \& Vitiello, A. (2021).
\textit{Implementing evolutionary optimization on actual quantum processors}.
Information Sciences, 575, 542-562.

\bibitem{srinivas1994adaptive}
Srinivas, M., \& Patnaik, L. M. (1994).
\textit{Adaptive probabilities of crossover and mutation in genetic algorithms}.
IEEE Transactions on Systems, Man, and Cybernetics, 24(4), 656--667.



\bibitem{goldberg1991comparative}
Goldberg, D. E., \& Deb, K. (1991).
\textit{A comparative analysis of selection schemes used in genetic algorithms}.
Foundations of Genetic Algorithms, 1, 69--93.







\bibitem{Buonaiuto2023bestpractices}
Buonaiuto, G., et al. (2023).
\textit{Best practices for portfolio optimization by quantum computing, experimented on real quantum devices}.
Scientific Reports, 13, 45392.    

\bibitem{Shunza2023discretepo}
Shunza, J., et al. (2023).
\textit{Application of quantum computing in discrete portfolio optimization}.
Journal of Digital Finance, 1(2), 1-10.    

\bibitem{Zaman2024poqa}
Zaman, K., et al. (2024).
\textit{PO-QA: A Framework for Portfolio Optimization using Quantum Algorithms}.
arXiv:2407.19857.

\bibitem{Naik2025financequantum}
Naik, A. S., \& al. (2025).
\textit{From portfolio optimization to quantum blockchain and security}.
Journal of Financial Innovation, 11, 25-40.

\bibitem{Gunjan2023quantuminspired}
Gunjan, A., Bhattacharyya, S., \& Hassanien, A. E. (2023).
\textit{Portfolio Optimization Using Quantum-Inspired Modified Genetic Algorithm}.
Smart Innovation, Systems and Technologies, 358, 665-673.

\bibitem{Haghighi2025eaqqa}
Haghighi, M. K., et al. (2025).
\textit{EAQGA: A Quantum-Enhanced Genetic Algorithm with Novel Entanglement-Aware Crossovers}.
arXiv:2504.17923.

\bibitem{nielsen}
Nielsen, M. A., Chuang, I. L. (2010). Quantum Computation and Quantum Information. Cambridge University Press.

\bibitem{combarro}
Combarro, E. F., González-Castillo, S., Di Meglio, A. (2023). A Practical Guide to Quantum Machine Learning and Quantum Optimization. Packt Publishing.





\bibitem{markowitz1952portfolio}
Markowitz, H. (1952). 
\textit{Portfolio selection}. 
The Journal of Finance, 7(1), 77--91.



\end{thebibliography}

\end{document}